\documentclass[main]{sshArticle}

\usepackage{stfloats}
\usepackage{nicefrac}
\usepackage{soul}
\definecolor{colorhbar}{RGB}{154, 181, 255}

\usepackage{amsmath}
\usepackage{amssymb}
\usepackage{graphicx}
\usepackage{gensymb}
\usepackage{textcomp}
\usepackage{bm}
\usepackage{placeins}
\usepackage{soul}
\usepackage{xcolor}

\title{Liquid Helium Cryogenic TEM below 1 \r{A}}

\author[1,2*]{Suk~Hyun~Sung}
\author[3,4*]{Maya~Gates}
\author[2*]{Nishkarsh~Agarwal}
\author[1]{Yang~Zhang}
\author[2]{William~Millsaps}
\author[2]{Miti~Shah}
\author[3]{Emily~Rennich}
\author[5]{\\Cong~Li}
\author[5]{Pu~Yu}
\author[6,7]{Miaofang~Chi}
\author[4,8]{Benjamin~H.~Savitzky}
\author[1†]{Ismail~El~Baggari}
\author[2,9†]{Robert~Hovden}

\affil[1]{The Rowland Institute at Harvard, Harvard University, Cambridge, MA 02138, United States}
\affil[2]{Department of Materials Science and Engineering, University of Michigan, Ann Arbor, MI 48109, United States}
\affil[3]{Department of Mechanical Engineering, University of Michigan, Ann Arbor, MI 48109, United States}
\affil[4]{h-Bar Instruments, Ann Arbor, MI 48103, United States}

\affil[5]{State Key Laboratory of Low Dimensional Quantum Physics and Department of Physics, Tsinghua University, Beijing, 100084, China}
\affil[6]{Center for Nanophase Materials Sciences, Oak Ridge National Laboratory, Oak Ridge, TN, United States}
\affil[7]{Thomas Lord Department of Mechanical Engineering \& Materials Science, Duke University, Durham, North Carolina 27708, United States}
\affil[8]{Department of Chemistry, Brown University, Providence RI, 02912, United States}
\affil[9]{Department of Physics, University of Michigan, Ann Arbor, MI, 48105, United States}
\affil[*]{Contributed equally}
\affil[†]{e-mail: ielbaggari@fas.harvard.edu, hovden@umich.edu}

\date{\today}

\firstAuthorLastName{Sung}

\abstract{
Next-generation cryogenic transmission electron microscopes (TEM) aim to achieve high-resolution imaging at ultracold sample temperatures (\textless~90~K) and over extended hold times. 
Lower temperatures enable atomic-scale characterization with improved beam and dose resilience for organic specimens and access to emergent electronic phases in quantum materials.
Side-entry liquid helium cooling stages presently lack the mechanical and thermal stability required to support sub-Angstrom information transfer in modern TEM.
Here we demonstrate sub-Angstrom atomic imaging TEM with a side-entry stage at specimen temperatures down to $\sim$20~K and with low stage drift and stable hold times.
}

\begin{document}

\maketitle
1
Liquid nitrogen cryogenic transmission electron microscopy (cryo-TEM) has driven atomic‐scale breakthroughs in structural biology~\cite{kuhlbrandt2014resolution, yip2020atomic} and, more recently, in materials science~\cite{li2017atomic,zachman2018cryo,elbaggari2018,zhu2023formation}. 
With liquid nitrogen cooling, sample temperatures in TEM typically reach 90--100 K~\cite{miaofang2024temp, schnitzer2025PRX}.
Next-generation cryo-TEM aims to reach lower temperatures using liquid helium cryogenics, to increase radiation protection \cite{glaeser1971limitations,siegel1974influence,chiu1986cryoprotection,iancu2006comparison,pfeil2019comparative,naydenova2022reduction} and characterize electronic phase transitions in quantum materials~\cite{matricardi1967electron,harada1992real,zhu2021cryogenic}. 
Long hold times, low drift, and atomic resolution at these ultracold temperatures are required to enable such experiments. 
However, existing cooling methods utilize small cryogenic dewars and are limited by rapid cryogen boil‐off, mechanical vibrations, and thermal drift, which prevent sustained sub‐Angstrom imaging~\cite{mun2024atomic}.
Liquid helium is sixty times more volatile than liquid nitrogen~\cite{wilks1967properties}, exacerbating the mechanical and thermal instabilities pervasive in cryo-TEM. 
In structural biology, Henderson et al. advocate that “the use of a microscope capable of cooling the specimen to liquid helium temperature should give more information from each micrograph”~\cite{henderson2004realizing}.
Russo's group recently solved the long-standing specimen movement problem in liquid helium cryo-TEM, using novel Au grids and illumination schemes~\cite{russo2025augrid}, heralding renewed interest in novel biological structure determination through ultracold electron microscopy.

Early TEM helium stage designs integrated cryogenic modules within the electromagnetic lenses~\cite{venables1963liquid,matricardi1967electron,fujiyoshi1991development,jiang2008backbone,borrnert2019dresden}. 
While offering low base temperatures, this architecture is incompatible with modern aberration-corrected electron microscopes that prioritize side-entry holders and narrow electromagnetic pole pieces for high resolution. 
For side-entry sample loading, vibrations are ruinous for resolution---structure blurs in real space and high-frequency structural peaks disappear in reciprocal space. 
In 2007, Klie and colleagues captured atomic resolution images using a side-entry stage during fleeting moments of stability~\cite{klie2007PRL}. 
Zhu's group has shown that stable imaging is generally only possible after the cryogen has fully evaporated—typically within 20 minutes—since the mechanical vibrations induced during active cooling are too severe~\cite{zhao2018direct,mun2024atomic}. 
However, once the cryogen is depleted, the sample temperature quickly rises and drift becomes severe~\cite{mun2024atomic}. 
A recent alternative design using continuous liquid helium flow achieves hold times exceeding 8 hours with high temperature stability, enabling TEM information transfer of 2.3~\AA{}~\cite{rennich2024ultra}.
In that initial implementation, information transfer below 1~\AA{} at ultracold cryogenic temperatures had not yet been demonstrated.

\begin{figure}[b!]
\centering 
  \includegraphics[width=0.97\linewidth]{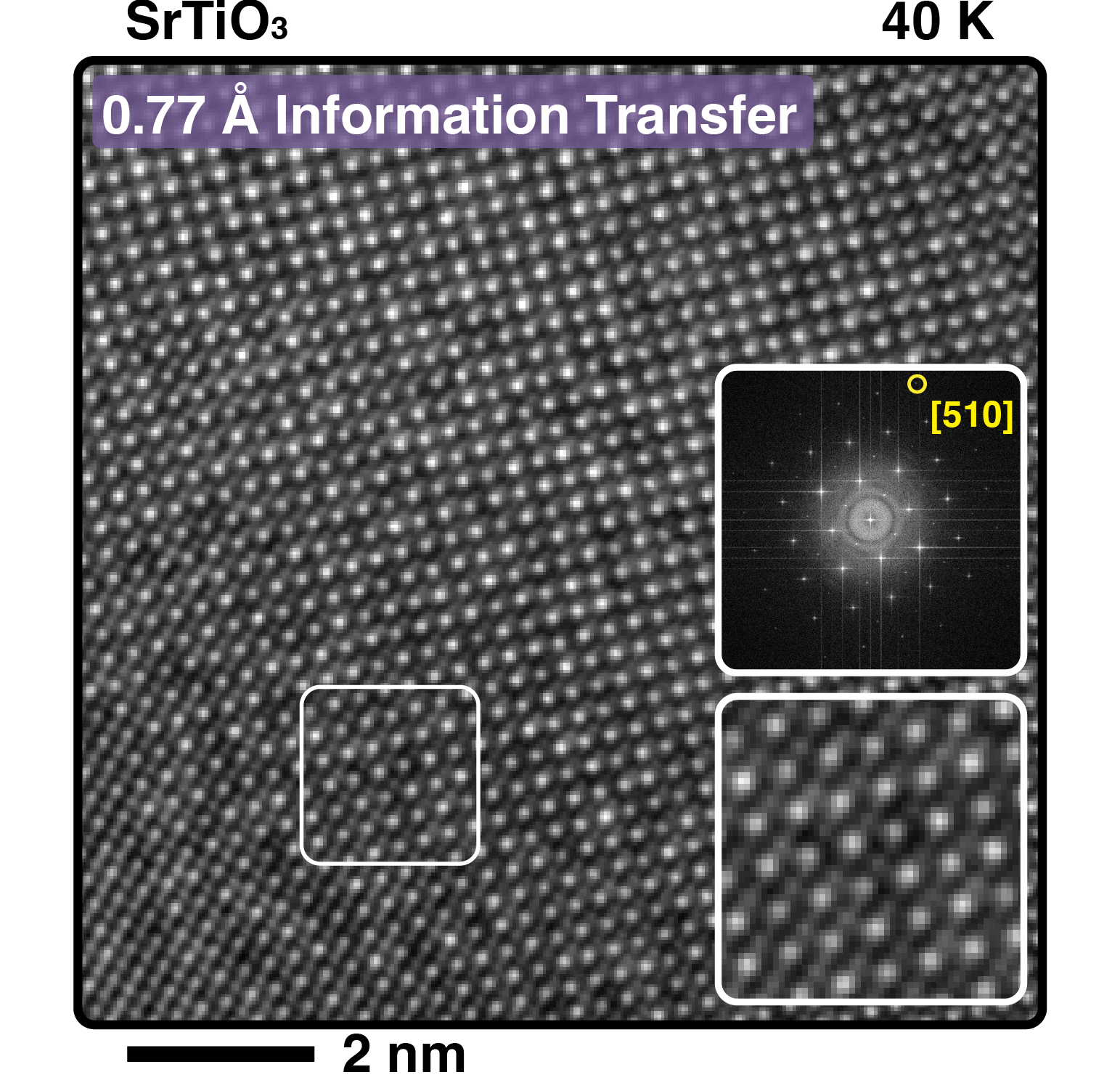}
  \caption{\textbf{Sub-Angstrom information transfer with ultracold cryogenic TEM imaging.} Atomic resolution BF-TEM images of SrTiO\textsubscript{3} taken at 40~K. Inset) Fourier transform of the image shows [510] Bragg peak corresponding to information transfer up to 0.77~Å.}
  \label{F:Fig1}
\end{figure}

Here we report atomic resolution imaging with sub-Angstrom information transfer, using a liquid‐helium TEM sample holder for modern aberration‐corrected transmission electron microscopes. 
The holder uses a continuous flow of liquid helium through a heat exchanger that thermally connects to the sample~\cite{rennich2024ultra}.  
TEM imaging with sub‐Angstrom information transfer is achieved by leveraging passive vibration isolation of the sample.
Through precise temperature regulation the stage achieves low drift---roughly twice lower than commonly used liquid nitrogen stages~\cite{goodge2020atomic}. 
Further, these stable imaging and cryogenic cooling conditions are maintained over extended periods (more than 24 hours), enabling extended data collection. The same stability is also achieved at intermediate temperatures, enabling imaging of structural changes across variable temperatures.
This ultracold specimen holder enables uninterrupted atomic‐resolution imaging with wide applicability to beam‐sensitive matter and electronic materials.

Figure~\ref{F:Fig1}a shows the atomic structure of a SrTiO\textsubscript{3} membrane taken at 40~K using the ultracold TEM specimen holder.
The holder was inserted into a Thermo Fisher Scientific Spectra 300 operating at 300~kV. 
Bright-field TEM (BF-TEM) images were acquired at 30~ms per frame and aligned using image registration~\cite{savitzky2018image}. 
The Fourier transform shows information transfer down to 0.77~\AA{}.
The sample temperature was tuned to 40~K where a quantum paraelectric phase emerges in SrTiO\textsubscript{3}~\cite{kustov2020STO}.

\begin{figure}[!t]
\centering
   \includegraphics[width=1.0\linewidth]{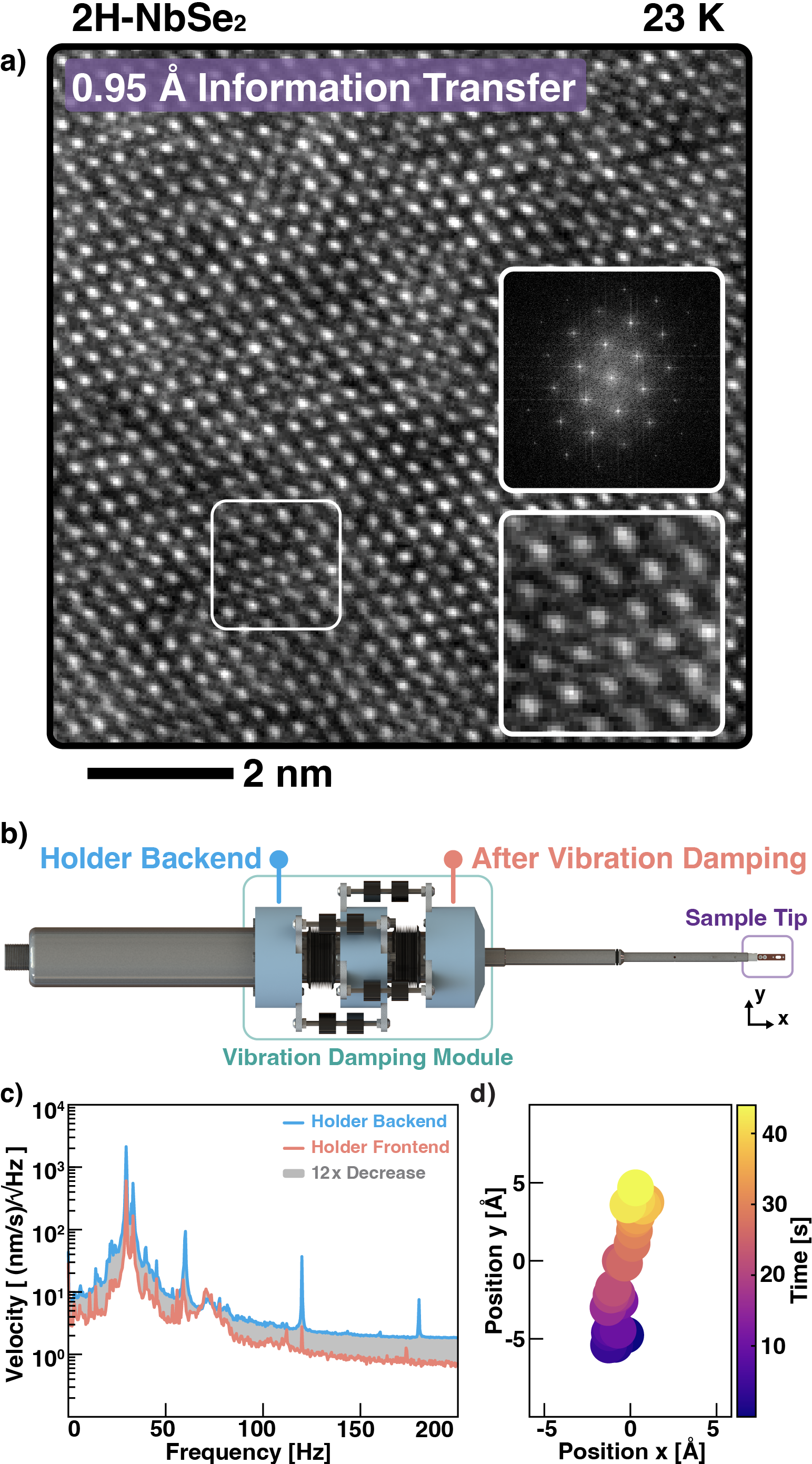}
  \caption{\textbf{Imaging stability of side-entry liquid helium cryo-TEM.} a) Atomic resolution BF-TEM image of 2H-NbSe\textsubscript{2} with 0.95~Å information transfer demonstrated at 23~K. 
  b) Vibration isolation is critical for achieving sub-Angstrom information transfer TEM imaging. A vibration damping module isolates incoming mechanical instabilities at holder backend from the sample tip.
  c) Spectral analysis of vibration measurements on the holder shows $\sim$12~times reduction in holder vibration. 
  d) Drift stability of 0.37~Å/s is achieved through millikelvin temperature stability. 
  Both vibrational stability and drift stability are critical for achieving sub-Angstrom information transfer.}
  \label{F:Fig2}
\end{figure}

Figure~\ref{F:Fig2}a shows a BF-TEM image of the atomic structure of two-dimensional 2H-NbSe\textsubscript{2} at 23 K, closer to the sample holder's base temperature.
Atomic-scale stability is enabled by two stage vibration isolation separating the specimen from the liquid helium flow in the holder backend (Fig.~\ref{F:Fig2}b).
The vibration isolator includes flexible high-vacuum bellows and compressed viscoelastic material for damping.
Figure~\ref{F:Fig2}c shows velocity spectra acquired using geophone vibration sensors placed before and after the isolation assembly; 
vibrations are damped by $\sim$12~times, thus enabling TEM imaging at atomic resolution.

Beyond vibration, sample drift due to unregulated temperature changes is a persistent limitation in cryo-TEM, causing resolution loss, image distortion, or the sample to shift from the field of view.
Figure~\ref{F:Fig2}d shows low sample drift, approximately 0.37~\r{A}/s. 
Stage drift at cryogenic temperature is due to unregulated temperature changes, from cryogenic boil-off or additional heat loads, leading to thermal expansion and contraction of the sample holder~\cite{goodge2020atomic,savitzky2018image}. 
As a point of comparison, high-resolution imaging in conventional side-entry liquid nitrogen cryo-TEM is performed under drift rates $\sim$0.6--1~Å/s range~\cite{elbaggari2018,goodge2020atomic}.
For the best imaging resolution and signal-to-noise ratio, cryo-TEM typically leverages fast multi-frame acquisition and image alignment to overcome sample drift and noise in single frames \cite{li2013electron,savitzky2018image}.

The atomic-scale stability and cryogenic cooling conditions are sustained for extended periods (over 24 hours), as long as helium is supplied from a large (60+ liters) external dewar.  
Previously, cryo-TEM experiments were limited by ultra-short hold times of liquid helium—often just tens of minutes. 
This vibration-damped continuous flow design now allows dramatically extended hold times in addition to atomic resolution, enabling more complex and longer in situ experiments, such as imaging across variable temperatures.

Stable atomic imaging at ultracold temperatures is a longstanding grand challenge in cryo-TEM for structural biology and materials science~\cite{minor2019cryogenic,russo2025augrid}.
This work demonstrates sub-Angstrom information transfer in a modern TEM at ultracold temperatures, using a vibration-isolated liquid-helium-cooled side-entry sample holder.
Ultracold temperatures below 90 K in state-of-the-art TEM are expected to improve the radiation tolerance, and hence resolution, in imaging biological materials. 
Recent work confirms a 1.5-fold improvement in radiation tolerance of mouse heavy-chain apoferritin with liquid helium cooling \cite{russo2025augrid}, motivating the further development of ultracold cryogenic TEM.
The same enhanced resilience may apply to other electron-beam-sensitive materials such as halide perovskites, organic crystals, and organic-inorganic interfaces.
Atomic-resolution imaging at variable cryogenic temperatures below 90 K is poised to reveal the structure across phase transitions in quantum materials.

{\printbibliography}

\section*{Methods}

\subsection*{Cryogenic TEM Experiment}
The h-Bar Instruments ULT cryo-TEM specimen holder uses continuous flow liquid helium cooling with internal cryogenic components operating under high vacuum conditions ($\sim$1.0$\times$10\textsuperscript{-7} torr). 
The holder has two-stage vibration isolation between the back-end heat exchanger and the front-end specimen stage.
Temperature was maintained using the h-Bar Instruments \textit{Mili} control box. 
In situ cooldowns were performed on a Thermo Fisher Scientific (TFS) Spectra 300 STEM/TEM operated at 300~kV. 
The microscope column pressure of $\sim$7.5$\times$10\textsuperscript{-8} torr was maintained at the specimen and cryogenic stage.
TEM images were acquired on a TFS Ceta-S CMOS camera with 30~ms (Fig.~\ref{F:Fig1}) and 25~ms (Fig.~\ref{F:Fig2}a) acquisition times per frame. 
Frames were aligned using a rigid registration algorithm~\cite{savitzky2018image}, then averaged over 94 (Fig.~\ref{F:Fig1}) and 20 (Fig.~\ref{F:Fig2}a) frames.

\subsection*{Sample Preparation}
SrTiO\textsubscript{3} thin films were deposited on acid-dissolving SrCoO\textsubscript{2.5} grown on LSAT(001) using pulsed laser deposition (PLD) with KrF laser ($\lambda=248~\textrm{nm}$, energy density of 0.9~J$\cdot$cm\textsuperscript{--2})~\cite{peng2022OxideBuffer}. 
SrCoO\textsubscript{2.5} sacrificial layer was grown at 730°C with repetition rate of 3~Hz; SrTiO\textsubscript{3} film was grown at 700°C with repetition rate of 2~Hz. 
All growths were performed at a constant oxygen partial pressure of 13~Pa. 
After growth, the samples were cooled to room temperature at a rate of 10°C$\cdot$min\textsuperscript{-1}. 
The films were then placed face-down onto TEM grids and immersed in 36\% acetic acid at room temperature until the SrCoO\textsubscript{2.5} sacrificial layer was completely dissolved, leaving freestanding SrTiO\textsubscript{3} membranes supported on the grids.

2H-NbSe\textsubscript{2} flakes were exfoliated from a single crystal onto a polydimethylsiloxane (PDMS) gel stamp and mechanically transferred on SiN\textsubscript{x} TEM grids using a home-built transfer system.

\section*{Acknowledgements}

Cryo-TEM experiments with sub-Angstrom imaging of 2H-NbSe\textsubscript{2} were performed in July 2024. 
Sub-Angstrom imaging of SrTiO\textsubscript{3} was performed in April 2025.
I.E., S.H.S. and Y.Z. acknowledge support from the Rowland Institute at Harvard.
R.H. acknowledges support from the U.S. Department of Energy, Basic Energy Sciences, under award DE-SC0024147.
M.G. and B.H.S. recognize support from the National Science Foundation under Small Business Innovation Research award 2322155. 
Electron microscopy experiments were conducted using the Michigan Center for Materials Characterization ((MC)\textsuperscript{2}) at the University of Michigan, with assistance from Tao Ma and Robert Kerns. 
We acknowledge Winfield Hill, Erik Madsen, Kal Banger, Alan Stern and Chris Stokes from the Rowland Institute at Harvard for their help with machining, electronics design, and cryogenics.

\section*{Author contributions}
I.E., R.H., S.H.S., N.A., B.H.S., W.M., M.S., M.G., M.C., Y.Z. performed cooldown experiments and in situ microscopy measurements.
I.E., R.H. S.H.S., N.A., M.G., E.J.R., designed and built the hardware. 
S.H.S. and N.A. optimized temperature control.
C.L. and P.Y. synthesized samples.
S.H.S., R.H., and I.E. prepared the manuscript. 
All authors reviewed and edited the manuscript.

\section*{Competing interests}
R.H., E.J.R. and I.E. are inventors on a filed patent application related to this work---the patent is owned by the University of Michigan and Harvard University.
R.H., I.E., B.H.S. and M.G. are involved in developing cryogenic electron microscopy technologies at h-Bar Instruments.
The remaining authors declare no competing interests.

\end{document}